\documentclass[twocolumn]{aastex631} 
\usepackage{csvsimple}
\usepackage{multirow,bigdelim}
\usepackage{amsmath}
\usepackage{eufrak}

\newcommand{\kms}{\mbox{km s$^{-1}$}}

\usepackage{colortbl}
\definecolor{tablegray}{rgb}{0.89, 0.89, 0.89}

\begin{document}

\title{X-rays Trace the Volatile Content of Interstellar Objects}
\correspondingauthor{Samuel H. C. Cabot}

\email{sam.cabot@yale.edu}

\author[0000-0001-9749-6150]{Samuel H. C. Cabot}
\affil{Yale University, 52 Hillhouse, New Haven, CT 06511, USA}

\author[0000-0002-9279-4041]{Q. Daniel Wang}
\affil{Astronomy Department, University of Massachusetts, Amherst, MA 01003, USA}

\author[0000-0002-0726-6480]{Darryl Z. Seligman}
\altaffiliation{NSF Astronomy and Astrophysics Postdoctoral Fellow}
\affil{Department of Astronomy and Carl Sagan Institute, Cornell University, 122 Sciences Drive, Ithaca, NY, 14853, USA}

\begin{abstract}
The non-detection of a coma surrounding 1I/`Oumuamua, the first discovered interstellar object (ISO), has prompted a variety of hypotheses to explain its {nongravitational} acceleration. Given that forthcoming surveys are poised to identify analogues of this enigmatic object, it is prudent to devise alternative approaches to characterization. In this study, we posit X-ray spectroscopy as a surprisingly effective {probe of volatile ISO compositions}. Heavily ionized metals in the solar wind interact with outgassed neutrals and emit high-energy photons in a process known as charge exchange, {and charge exchange induced X-rays from comets and planetary bodies have been observed extensively in our Solar System}. We develop a model to predict the X-ray flux of an ISO based on its chemical inventory and ephemeris. {We find that} {while standard cometary constituents, such as H$_2$O, CO$_2$, CO, and dust are best probed via optical or infrared observations, we predict strong X-ray emission generated by charge exchange with extended comae of H$_2$ and N$_2$ --- species which lack strong infrared fluorescence transitions}. We find that {\it XMM-Newton} {would have been sensitive to charge exchange emission from} 1I/`Oumuamua {during the object's close approach to Earth}, and that constraints on composition may have been feasible. We argue for follow-up X-ray observations of newly discovered ISOs with close-in perihelia. Compositional constraints on the general ISO population could reconcile the apparently self-conflicting nature of 1I/`Oumuamua, and provide insight into the earliest stages of planet formation in extrasolar systems.

\end{abstract}

\section{Introduction}
Knowledge of the compositions of minor bodies has considerably influenced our understanding of the early Solar System. While investigations of their dynamical histories are limited by the Solar System's chaotic nature \citep{Wisdom1980,Laskar1989,Batygin2008,Laskar2009}, compositional measurements of minor bodies --- especially of their interiors --- have the potential to reveal their formation environments \citep{Oberg2011}. 

Remarkably, the Solar System likely ejected tens of earth masses of volatile rich material during the early stages of its formation and evolution \citep{Hahn1999,Gomes2004,Tsiganis2005,Morbidelli2005,Levison2008,Raymond2018b,Raymond2020}. Therefore, it is reasonable to expect that extrasolar planetary systems contribute to a galactic population of interstellar comets, and that some will encounter the Solar System on hyperbolic trajectories \citep{Moro2009, Engelhardt2014,Cook2016}. 
Naturally, it came as a surprise that 1I/`Oumuamua, the first discovered ISO, exhibited none of the typical properties of Solar System comets \citep[for recent reviews,  see][]{MoroMartin2022,Jewitt2022}.
It had no visible coma \citep{Meech2017,Jewitt2017,trilling2018spitzer,Bolin2018}, an extreme shape \citep[e.g.,][]{Knight2017,Mashchenko2019}, a reddened reflection spectrum \citep{masiero2017spectrum}, a young dynamical age \citep[e.g.,][]{mamajek2017,Hsieh2021} and {non-zero} {nongravitational} acceleration \citep{Micheli2018}.

The second discovered ISO, 2I/Borisov, was unambiguously a comet
\citep{Jewitt2019}, 
perhaps more inline with expectations. Its cometary activity was readily measured and typical carbon and nitrogen bearing species were detected \citep{Opitom:2019-borisov, Kareta:2019, lin2020,Bannister2020,Xing2020,Aravind2021}. {\it ALMA} and {\it Hubble} observations revealed that it was enriched in CO relative to H$_2$O \citep{Bodewits2020, Cordiner2020}. This finding {suggests} that 2I/Borisov formed exterior to the CO snowline in {the} protoplanetary disk \citep{Price2021,Seligman2022} of a {very young}  \citep[][]{Lisse2022} or a very carbon-enriched system \citep[cf.][]{Bodewits2020}.

`Oumuamua's former provenance remains in question, and dynamically backtracing either ISO to their respective host systems is {difficult} \citep{mamajek2017, BailerJones2018}. 1I/`Oumuamua's peculiar properties and elusiveness to follow-up spectroscopy have prompted a variety of theories regarding its origin. If the measured {nongravitational} acceleration was driven by cometary outgassing, this would be energetically consistent with a composition of H$_2$ \citep{fuglistaler2018solid,Seligman2020h2,Levine2021_h2}, N$_2$ \citep{Jackson2021,Desch2021}, or CO \citep{Seligman2021}. Confirming an H$_2$ or N$_2$ composition would have immediately revealed a new mechanism of generating minor bodies: formation in the cores of molecular clouds \citep{Seligman2020h2}, or ejection of ice from the surfaces of Pluto analogues due to impact events \citep{Desch2021}.  Alternatively, if the acceleration was caused by radiation pressure \citep{Micheli2018}, then 1I/`Oumuamua must have been extremely porous \citep{moro2019fractal,Sekanina2019b,Flekkoy2019,luu2020oumuamua} or thin \citep{bialy2018radiation}. {While H$_2$O, CO$_2$, and CO species in cometary comae are easily measured via infrared spectroscopy, the possible presence of difficult to detect homonuclear H$_2$ or N$_2$ prompts a search for alternative methods for characterizing interstellar comets}.

This study presents {such} a new approach: X-ray observations of solar wind (SW) charge exchange (CX). X-ray observations of an interstellar object should immediately reveal the presence of an outgassed coma. CX involves the transfer of electrons from a cool neutral medium to heavily ionized metals in the SW. Following the unexpected detection of X-rays emanating from {C/1996 B2 (Hyakutake)} \citep{Lisse1996, Cravens1997}, CX has been observed with a multitude of other comets \citep{Dennerl1997, Cravens2000b, Cravens2002, Lisse1999b, Lisse2004, Bodewits2007}. CX conveys information about the composition and speed of the {solar wind} \citep{Beiersdorfer2001, Bodewits2007, Gu2016}. Studies have also linked CX emission to physical and chemical properties of the minor body and its coma through the morphology of the X-ray emission profile \citep{Wegmann2004} and emission line ratios \citep{Mullen2017}.

{Indeed,} there are encouraging prospects for discovering additional ISOs in the coming decade.
Population synthesis models of the galactic population of ISOs have demonstrated that the Rubin Observatory Legacy Survey of Space and Time (LSST) will detect $1-2$ 1I/`Oumuamua analogues every year \citep{Hoover2022}. 

This paper is organized as follows. {In \S\ref{sec:motive}, we discuss the advantages and limitations of X-ray observations for characterizing minor bodies, and outline a simple model of charge exchange emission.} We apply our model to 1I/`Oumuamua and 2I/Borisov in \S\ref{sec:consider}, and compare our predictions to the known correlation between X-ray and visible-band flux. Our model suggests that if 1I/`Oumuamua did indeed outgas a coma, that CX would have been detectable with current-generation X-ray facilities during the period that {it approached Earth}. \S\ref{sec:expect} establishes expectations for detecting CX with new ISOs discovered by {the} {Rubin Observatory}. Finally, our conclusions are summarized in \S\ref{sec:con}. {We refer the reader to Appendix~\ref{sec:review} for a brief review of CX fundamentals, and to Appendix~\ref{sec:coma} for the details of our CX model.}

\section{Motivation \& Model}
\label{sec:motive}

{This section establishes use cases for X-ray observations of interstellar objects, and sets forth an analytic model of X-ray emission from outgassed comae. It also specifies the traits of ISOs (i.e., bulk composition and orbital trajectory) from which we expect robust detection of X-ray emission with present-day facilities.} 

\subsection{Information Gained from X-ray Observations}

{As a primer, it is useful to review the prototypical case of CX from an outgassed coma:} {C/1996 B2 (Hyakutake)}, a long-period comet which came within 0.23 au of the Sun at perihelion. {Infrared observations yielded unambiguous detections of H$_2$O, CO, CH$_4$, and C$_2$H$_6$ \citep{Mumma1996, Russo2002}, Contemporaneously,} significant X-ray emission was detected out to $\sim100,000$ km (6.7$\times10^{-4}$ au) from its nucleus by \citet{Lisse1996}. CX was suggested as the responsible process by \citet{Cravens1997}, and reaffirmed by \citet{Wegmann1998}. Other mechanisms (e.g., thermal bremsstrahlung, scattering, and fluorescence) may be ruled out \citep{Krasnopolsky1997, Lisse1999b, Lisse2004}. {Since this detection,} sophisticated models of CX with cometary comae have been developed, such as the hydrodynamical simulations by \citet{Wegmann2004}, {and CX has been observed with comae of numerous other comets \citep[e.g.][]{Ewing2013, Bodewits2007}.}

{X-rays have been conventionally of limited value for compositional studies of minor bodies. For one, CX detections arise only from significantly extended comae. Moreover, the most immediate insights conveyed are the morphology of the coma and the properties of the solar wind \citep{Lisse2004}. Infrared/radio spectroscopy remains the {\it de facto} technique for detecting molecular species which exhibit solar-pumped fluorescence transitions \citep{Crovisier1983, BockeleeMorvan2004}. Such species \citep[reviewed by e.g.][]{Biver2016, Bockelee2017} include H$_2$O, CO, CO$_2$, CH$_4$, C$_2$H$_2$, C$_2$H$_6$, and HCN. Complementary to infrared observations, UV spectroscopy can measure the production of some additional species \citep{Biver2022} such as noble gases and homonuclear diatomic molecules.} However, except for a small number of tenuous reports of their sublimation \citep[e.g., Ar in Hale-Bopp,][]{Stern2000} noble gases remained elusive in comets prior to {\it in situ} detections in 67P \citep{Rubin2018} by {the European Space Agency's} {\it Rosetta} {mission}. {{\it Rosetta} also confirmed the presence O$_2$ and N$_2$ in comets \citep{Rubin2015, Bieler2015}.}

{This study explores a novel use case for X-ray observations: to reveal an outgassed coma that is invisible in the infrared, and which may arise from the perihelion passage of an interstellar object. While the composition of  1I/`Oumuamua remains debated, a number of candidate species  (Table~\ref{tab:chem}) are monotomic or homonuclear diatomic, and will not fluoresce in the infrared. However,} CX will occur at a neutral medium, even one that does not {produce} observable {thermal emission, fluorescence, or scattering}. Therefore, the outgassed coma of an interstellar object may be X-ray bright even in the absence of infrared and optical detections. {In particular, X-ray observations offer a promising avenue for remote sensing an H$_2$ or N$_2$ coma. Both of these species are thermodynamically consistent with 1I/`Oumuamua's {nongravitational} acceleration under outgassing, and comprise the focus of this study. Per \citet{Seligman2021}, we benchmark results to a CO composition \citep[c.f.][]{trilling2018spitzer}; although this molecule is more readily detected in the infrared.}

\begin{table}
\centering
\caption{{Cryogenic} chemical properties for volatiles which may be constituents of ISOs. For each species, we report the temperature of sublimation ($T_{\rm Sub}$), ideal adiabatic index ($\gamma$), enthalpy of sublimation ($\Delta H$), and sound speed ($c_s$) per Equation~\ref{eqn:wchem}. We also report the total energy input to a single particle ($\mathcal{H}$) per Equation~\ref{eqn:enthalpyrelation}. Data from \citet{Shakeel2018}, except for CO which is obtained from \citet{Stephenson1987} and the NIST database.}
\medskip
\setlength{\tabcolsep}{4.0pt}
\begin{tabular}{@{}cccccc@{}}
\hline
Species & $\textit{T}_{\rm Sub}$ & $\gamma$ & $\Delta H$ & $c_s$ & $\mathcal{H}$ \\
 & ($\rm{K}$) & & (${\rm kJ \, mole^{-1}}$) & (km s$^{-1}$) & (${\rm kJ} \times 10^{-23}$)\\
\hline
H$_2$  & 6    & 7/5 & 1     & 0.19 & 0.18 \\
Ne     & 9    & 5/3 & 1.9   & 0.11 & 0.34 \\
N$_2$  & 25   & 7/5 & 7.34  & 0.14 & 1.27 \\
CO     & 58   & 7/5 & 7.6   & 0.15 & 1.37 \\
Ar     & 30   & 5/3 & 7.79  & 0.15 & 1.36 \\
O$_2$  & 30   & 7/5 & 9.26  & 0.10 & 1.60 \\
Kr     & 40   & 5/3 & 11.53 & 0.12 & 2.01 \\
Xe     & 55   & 5/3 & 15.79 & 0.12 & 2.75 \\
CO$_2$ & 82   & 8/6 & 28.84 & 0.14 & 4.94 \\
H$_2$O & 155  & 8/6 & 54.46 & 0.31 & 9.33 \\
\hline
\end{tabular}
\label{tab:chem}
\end{table}

\subsection{Scaling Relationships}
\label{sec:scalerel}

{In Appendix~\ref{sec:coma}, we develop a model that links X-ray flux to a minor body's physical properties.} Specifically, it depends on the following key parameters: the neutral production rate $(Q_{\rm gas})$, mean radius of the nucleus $(a)$, perihelion distance $(r_{\rm peri})$, sublimation + kinetic energy of the relevant volatiles $(\mathcal{H})$, geocentric distance $(\Delta_e)$, solar wind number density $(n_{\rm SW})$, CX interaction area of the coma $(\mathcal{A})$, and X-ray lumonisty $(L_{\rm X})$ and flux $(F_{\rm X})$. The pertinent relationships are as follows:

\begin{align}
    Q_{\rm gas} &\propto a^2 r_{\rm peri}^{-2} \mathcal{H}^{-1} \\
    \mathcal{A} &\propto Q_{\rm gas}^2 {\rm \, (to \, leading \, order)} \\
    n_{\rm SW} &\propto r_{\rm peri}^{-2} \\
    L_{\rm X} &\propto n_{\rm SW}\mathcal{A} \\
    F_{\rm X} &\propto \Delta_e^{-2}L_{\rm X} \propto a^4r_{\rm peri}^{-6}\mathcal{H}^{-2} \Delta_e^{-2}.
    \label{eqn:prop}
\end{align}

\begin{figure*}
\begin{center}
\includegraphics[width=\linewidth,angle=0]{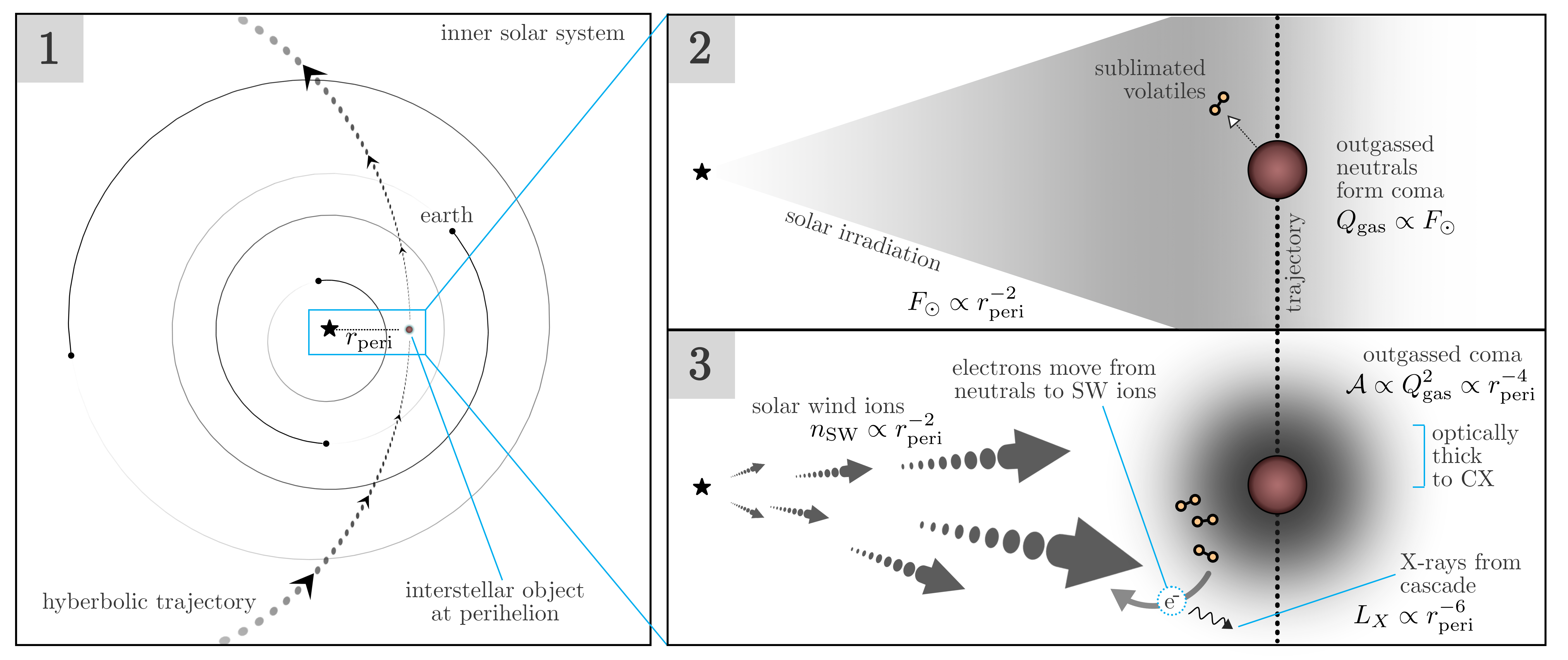}
\caption{Schematic of solar wind charge exchange with the outgassed neutrals in the coma of an ISO. {\it Panel 1}: An instantaneous view of the inner solar system and the trajectory of an ISO, currently at perihelion ($r_{\rm peri}$). {\it Panel 2}: The flux of solar radiation ($F_{\odot}$) heats the volatiles on the ISO's surface, which causes sublimation and outgassing of neutral volatiles at a rate $Q_{\rm gas}$. {\it Panel 3}: Solar wind ions (with number density $n_{\rm SW}$) interact with the outgassed coma through charge exchange. Electrons are transferred from the neutral coma to the bare or H-like {solar wind} ions, and the subsequent cascade emits X-rays. The coma's interaction cross-section ($\mathcal{A}$) scales with $Q_{\rm gas}^2$. As a result, the X-ray luminosity follows $L_{\rm X} \propto r_{\rm peri}^{-6}$ {(Equation~\ref{eqn:prop}, \S\ref{sec:scalerel})}.}
\label{fig:schematic}
\end{center}
\end{figure*}

{In our model, constants of proportionality are calibrated to measurements of C/1996 B2 (Hyakutake) (\citet{Lisse1996}, Appendix~\ref{sec:coma})}. The above scaling relationships demonstrate that the X-ray brightness of a minor body can be significantly enhanced by a close-in perihelion passage and a particularly volatile composition. {Figure~\ref{fig:schematic} shows a schematic of such an ISO encounter, where the scaling relationships and underlying physical processes are depicted in the individual panels.}

\subsection{Detection Thresholds}

In this section we guage the feasibility of detecting CX with an outgassed coma, including the {requisite} X-ray flux and exposure durations. Our predictions are for the EPIC pn detector onboard {\it XMM-Newton}, which has the advantage of a larger effective area (approximately $A = 900$ cm$^2$ {at 0.5 kev}) compared to the other space-based X-ray detectors. To estimate the detectability of a comet or ISO, one needs to account for the noise contribution from the background, which is chiefly made of background AGNs and the diffuse hot plasma emission from the general interstellar and circumgalactic media of the Galaxy, the Local Bubble, {and} solar wind charge exchange emission within the heliosphere \citep[e.g.,][]{McCammon2002, Koutroumpa2012}. {Based on archival blank-sky EPIC pn observations,} we estimate that the total background intensity is $\Sigma = 1.4\times10^{-3}$ {counts}
s$^{-1}$ arcmin$^{-2}$ in the $0.5-0.7$ keV band, for example, which encloses the typically strong O{\sc vii} and O{\sc viii} K$\alpha$ lines. {The total background intensity also includes the contribution from non-X-ray (cosmic-ray induced instrument) events.} Let $\Phi_X$ denote the X-ray flux at the detector due to CX with the target (units of {counts} s$^{-1}$ cm$^{-2}$), and let $\tau$ denote the exposure duration. Furthermore, we assume that nearly all of the signal is contained within a solid angle $\Omega$ for simplicity. Then the signal-to-noise (S/N) of the CX signature is

\begin{equation} \label{eqn:snr}
    {\rm S/N} = \Phi_X A \sqrt{\tau} / \sqrt{\Phi_X A + \Sigma \Omega} \, .
\end{equation}
Alternatively, this equation {can also} be rearranged into an expression for $\tau$ for a desired S/N threshold. 

One caveat is that a continuous exposure may be ineffective for fast-moving targets. {Photon events may, however, be spatially calibrated across discrete slews that keep the object on the focal plane array (FPA). This enables {multiple on-target pointings and} reconstruction of events in the cometocentric frame based on the attitude of the telescope. The maximum exposure duration {for any one pointing is then} limited by the field of view of the telescope and the angular velocity of the target. For reference, 1I/`Oumuamua's angular velocity reached $1.4\times10^{-4}$ deg. s$^{-1}$ at closest approach to Earth. The crossing time for {\it XMM-Newton}'s {$30'\times30'$ field of view would have been up to $\sim 3.5$ ks}}.

Space-based X-ray observatories have minimum solar elongation angles ($\psi_{\odot}^{\rm min}$) for reasons pertaining to {detector safety (e.g., avoiding burn-out incidents similar to ROSAT's)}, thermal stability, and power supply. For {\it Chandra}, $\psi_{\odot}^{\rm min} = 46.4^{\circ}$, and for {\it XMM-Newton}, $\psi_{\odot}^{\rm min} = 70^{\circ}$. These constraints permit observations of small bodies at $R\geq0.72$ au and $R\geq0.94$ au, respectively. These lower bounds can only be attained under optimal alignment between Earth and the object. {Future missions {will likely} have similar safety exclusion zones to protect equipment at the FPA}.

\begin{figure}
\begin{center}
\includegraphics[width=\linewidth,angle=0]{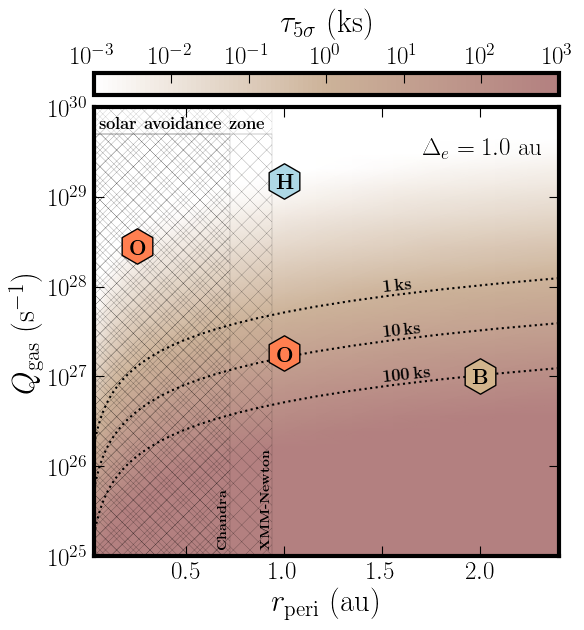}
\caption{Exposure time (ks) required for a 5$\sigma$ detection of CX X-ray emission (Equation~\ref{eqn:lxest}), as a function of outgassing rate $Q_{\rm gas}$ and perihelion distance $r_{\rm peri}$. Geocentric distance is fixed at $\Delta_e = 1$ au. The background rate and effective area correspond to that of EPIC pn on {\it XMM-Newton}. For reference, markers are plotted at locations representative of real minor bodies: light-blue (`H') corresponds to $Q_{\rm gas}=1.5\times 10^{29}$ s$^{-1}$ observed at 1.0 au --- similar to {C/1996 B2 (Hyakutake)}; tan (`B') corresponds to $Q_{\rm gas}=10^{27}$ s$^{-1}$ observed at 2.0 au --- similar to 2I/Borisov; coral (`O') corresponds to $Q_{\rm gas}=2.8\times 10^{28}$ s$^{-1}$ observed at 0.25 au {and $Q_{\rm gas}=1.8\times 10^{27}$ s$^{-1}$ observed at 1.0 au} --- {estimates} corresponding to 1I/`Oumuamua {under a CO composition}. {The solar elongation angles of {\it Chandra} ($\psi_{\odot}^{\rm min} = 46.4^{\circ}$) and {\it XMM-Newton} ($\psi_{\odot}^{\rm min} = 70^{\circ}$) impose lower bounds ($\sin\psi_{\odot}^{\rm min}$) on observable $r_p$, and their respective restricted zones are hatched in the figure. The lower bound is attained only in {the ideal orbital case and observable heliocentric distances are typically significantly larger due to spacecraft limitations.}}
}
\label{fig:Gq5sig}
\end{center}
\end{figure}

Figure~\ref{fig:Gq5sig} depicts the exposure times necessary to achieve S/N $= 5$ as a function of $Q_{\rm gas}$ and $r_{\rm peri}$, specifically for the {\it XMM-Newton} pn detector. For simplicity, the geocentric distance is assumed $\Delta_e = 1$ au. The X-ray luminosity of CX is given by Equation~\ref{eqn:lxest}, and the angular extent of the emission is set by $\tilde{b}$ (Equation~\ref{eqn:newb}). Markers in Figure~\ref{fig:Gq5sig} are located at outgassing rates and perihelia of real minor bodies (the outgassing rate of 1I/`Oumuamua was not directly measured, but is estimated in \S\ref{sec:consider}).

This figure immediately conveys the feasibility of detecting CX emission from outgassed comae with currently operational facilities. For modest exposure times of order $10$ ks, CX should be detectable when $Q > 10^{27}$ s$^{-1}$ over a range of $r_{\rm peri}$. Qualitatively, this finding is consistent with previous high-significance {\it ROSAT} detections of CX with six comets \citep{Dennerl1997}, whose ougassing rates were between $Q_{\rm gas}=2\times10^{27}-9\times10^{29}$ s$^{-1}$. The heliocentric distances in this sample were between $R = 0.99-1.96$ au, and geocentric distances between $\Delta_e = 0.12-1.61$ au (these comets are discussed further in \S\ref{sec:cxcomets}). While the minor bodies need not be observed at perihelion, our model indicates that $F_{\rm X}$ scales more strongly with {heliocentric distance} ($R$) than with $\Delta_e$ \citep{Lisse1999b} per Equations~\ref{eqn:production} \& \ref{eqn:nsw}. As discussed in the next section, 1I/`Oumuamua's {proximity to the Earth and Sun} would have made it amenable to X-ray observations. 
On the other hand, 2I/Borisov would have been a more difficult case because it only came within {a minimum} $2$ au of the Sun.

\section{Detectability of X-rays from 2I/Borisov and 1I/`Oumuamua} \label{sec:consider}

We proceed to estimate the X-ray luminosity for the only two identified ISOs {that traversed our Solar System}: 1I/`Oumuamua and 2I/Borisov. We first estimate $L_{\rm X}$ using the analytic framework presented in the previous section. We then make an independent estimate based on known trends between optical and X-ray luminosities for comets. 

We briefly review pertinent measurements of both ISOs. {From its photometric lightcurve structure, 1I/`Oumuamua is estimated to have had} an oblate 6:6:1 ellipsoidal geometry \citep{Mashchenko2019}. Its albedo ($p$) was not constrained, so its true dimensions are uncertain. \citet{Meech2017} {and \citet{Bolin2018} measured an effective radius of $\sim0.1$ km} assuming an albedo of 0.04. The light curve fit by \citet{Mashchenko2019} yielded dimensions $115 \times 111 \times 19$ m with $p=0.1$. For our order-of-magnitude calculations, we treat 1I/`Oumuamua as spherical with radius of $a=100$ m. However, we emphasize that the true dimensions are albedo-dependent, and the lengths scale as $1/\sqrt{p}$. At perihelion of $r_{\rm peri}=0.25$ au, 1I/`Oumuamua's distance to Earth was $\Delta_e = 1.23$ au with an apparent magnitude $V=21$. The object reached an apparent magnitude of $V=19.7$ at its closest approach to Earth with $\Delta_e=0.16$ au. Infrared {\it Spitzer} observations produced non-detections when 1I/`Oumuamua was outbound at $\sim2$ au, providing upper limits on the production rates of  CO or CO$_2$ \citep{trilling2018spitzer}. However, sporadic outgassing of CO might reconcile the {nongravitational} acceleration and {\it Spitzer} non-detection \citep{Seligman2021}. 

The nuclear radius of {comet-like} 2I/Borisov was constrained to $0.2 \, {\rm km}<a<0.5 \, {\rm km}$ based on its brightness profile and {nongravitational} acceleration assuming {a} range of plausible bulk densities \citep{Jewitt2019, Bolin2020}. We adopt a nominal radius of $a=300$ m. {Unlike 1I, gas emission
activity typical of comets was definitely detected from this body}. When contemporaneous measurements of H$_2$O and CO production rates were obtained after perihelion ($r_{\rm peri}=2$ au), CO dominated the outflow. Specifically \citet{Bodewits2020} measured $Q({\rm CO}) = 7.5\times10^{26}$ s$^{-1}$ at 3.3 days after perihelion and $Q({\rm CO}) = 1.07\times10^{27}$ s$^{-1}$ at 22.4 days after perihelion. Also, \citet{Cordiner2020} measured $Q({\rm CO}) = 4.4\times10^{26}$ s$^{-1}$ in observations conducted $7-8$ days after perihelion. While water production exceeded $Q({\rm H_2O}) = 10^{27}$ s$^{-1}$ prior to perihelion, it dropped below $6\times10^{26}$ s$^{-1}$ for observations conducted after perihelion \citep{Bodewits2020}. \citet{Jewitt2020} measured $V=16.6$ at 26 days after perihelion, when 2I/Borisov's distances were $R=2.1$ au and $\Delta_e = 1.9$ au.

In the following, $Q_{\rm gas}$ denotes the total outgassing rate across all neutral volatiles. 
In certain instances, a dominant species $X$ was determined, in which case its outgassing rate is denoted $Q({X})$. {Our predictions for $L_{\rm X}$ neglect subtle dependencies on properties of the solar wind and coma. In the optically thick regime, all {solar wind} minor ions will undergo charge exchange. However, the X-ray lines emitted depend on the neutral composition \citep{Mullen2016}. Also, CX cross sections depend strongly on the {solar wind} velocity \citep{Kharchenko2001, Beiersdorfer2001, Bodewits2007}.
}

\subsection{Analytic Predictions for 2I/Borisov}

For 2I/Borisov, we adopt the measured production rate $Q({\rm CO}) = 1.07 \times 10^{27}$ s$^{-1}$ at $R = 2.07$ au \citep{Bodewits2020}. We note that Equation~\ref{eqn:production} would predict $Q_{\rm gas} = 5.9\times 10^{26}$ s$^{-1}$ near perihelion, and modifying the scaling relation to take into account a CO composition (i.e., an $\mathcal{H}$ appropriate for CO ice) yields $Q_{\rm gas} = 4.0\times 10^{27}$ s$^{-1}$, which are both within an order-of-magnitude of the observed value. Accounting for the reduced solar wind density at 2 au, we calculate $L_{\rm X} = 7.5\times10^{11}$ erg s$^{-1}$ {(see Table~\ref{tab:lx})}. A 900 cm$^{2}$ effective area (characteristic of EPIC pn) at Earth would detect $6.3\times10^{-5}$ {counts} s$^{-1}$, which would require a considerable 400 ks exposure for a $5\sigma$ detection. Indeed, most previous CX detections in comets have involved much brighter X-ray luminosities $L_{\rm X} > 10^{14}$ erg s$^{-1}$ \citep{Lisse1999b, Lisse2004}.

\begin{figure*}
\begin{center}
\includegraphics[width=0.6\linewidth,angle=0]{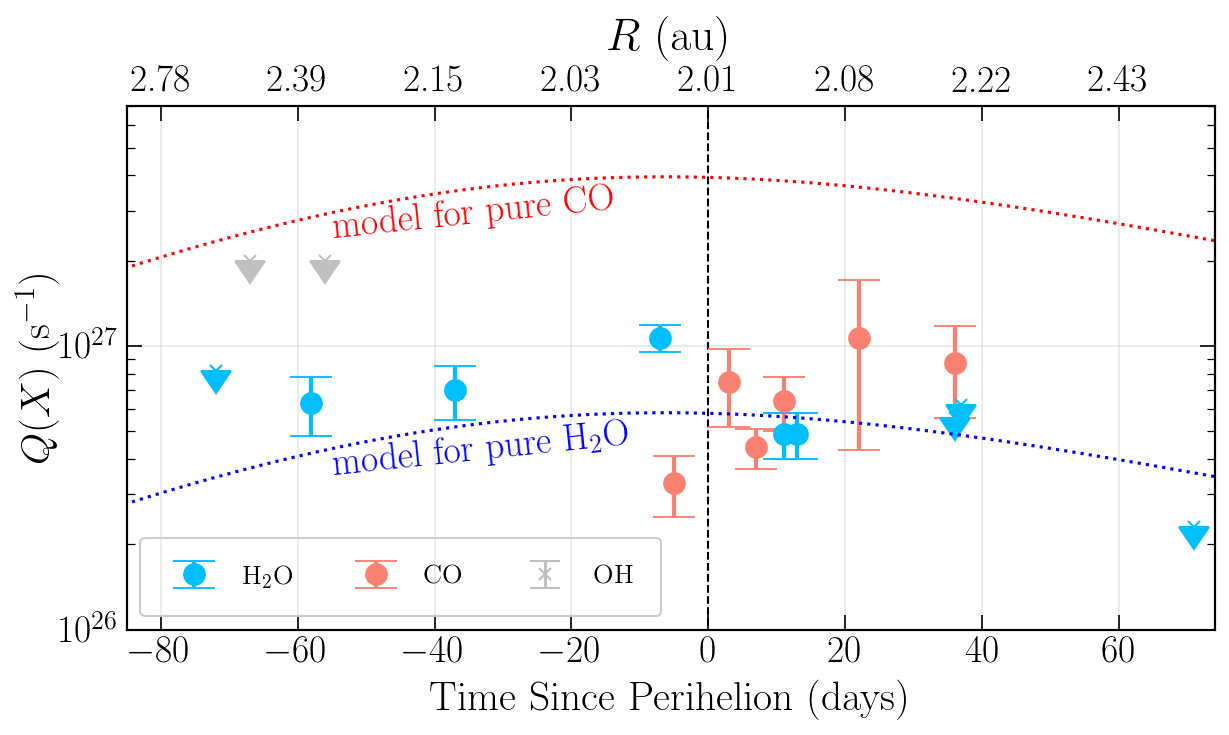}
\includegraphics[width=0.37\linewidth,angle=0]{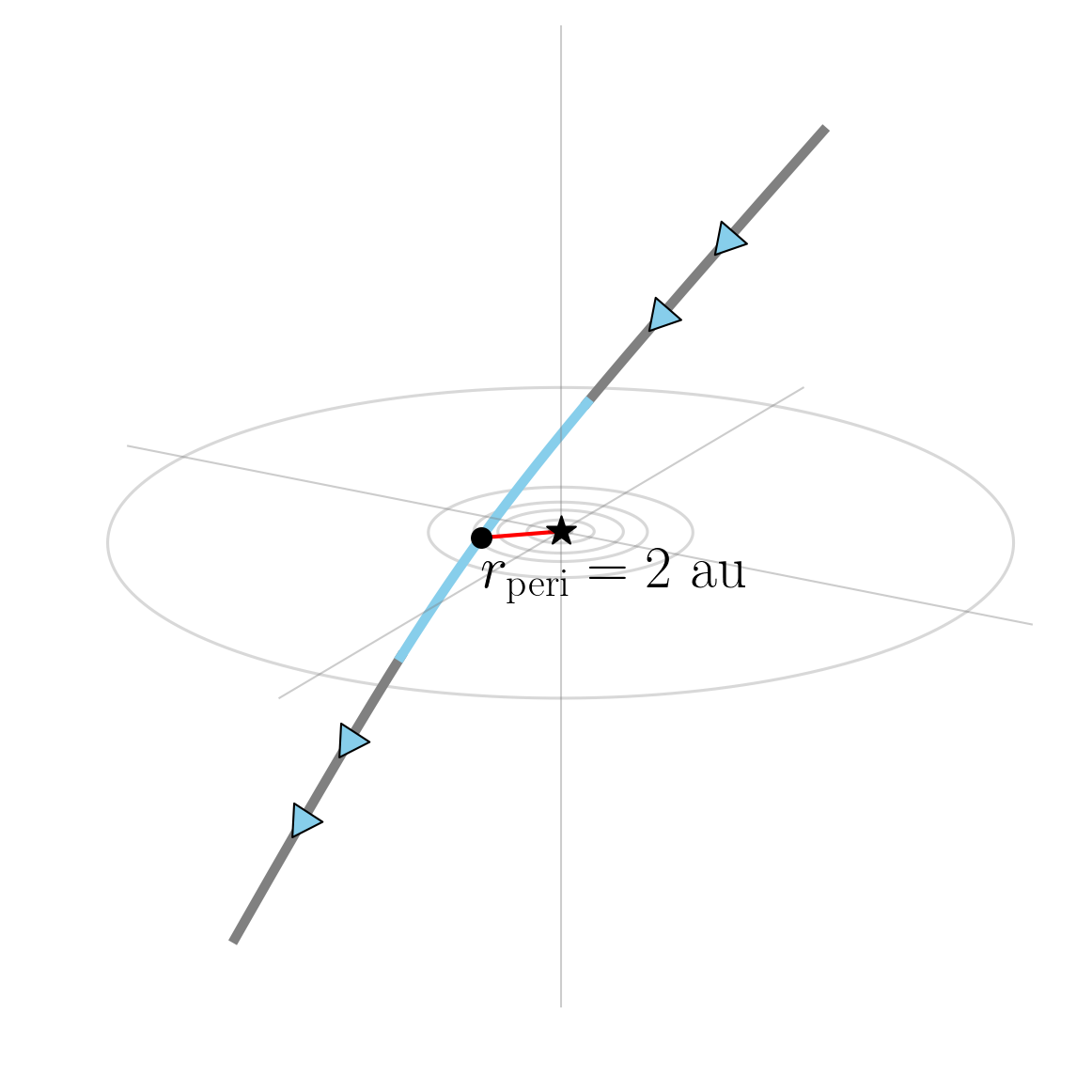}
\caption{Time-evolution of production rates for 2I/Borisov near perihelion (for CO, H$_2$O, and OH). Figure adapted from \citet{Seligman2022}. References for data are provided in their Table 2. The measurements demonstrate that CO production was substantially greater than H$_2$O production after perihelion --- the only time when both species were observed contemporaneously. Our production rate model (Equation~\ref{eqn:production}) is plotted as dashed lines, holding 2I/Borisov's radius as a constant $a=0.3$ km, and adopting an appropriate $\mathcal{H}$ for H$_2$O and CO. We obtained 2I/Borisov's heliocentric distance with a \texttt{rebound} \citep{Rein2012} simulation using the Mercurius integrator \citep{Rein2019}. For reference, the trajectory of 2I/Borisov through the Solar System is shown at right, with perihelion marked. The solid blue portion corresponds to the timespan plotted in the left panel.
}
\label{fig:borisovcomp}
\end{center}
\end{figure*}

Since there are observations of 2I/Borisov at multiple points along its trajectory, we may perform a more detailed comparison between our model and measurements of outgassing over time.
We plot the observed time-evolution of $Q({\rm H_2O})$ and $Q({\rm CO})$ in Figure~\ref{fig:borisovcomp} with our production rate model (Equation~\ref{eqn:production}). Our model, which was calibrated to observations of {C/1996 B2 (Hyakutake)}, predicts {water} production rates within {a factor a few} of those observed. {The discrepancies may be due to composition inhomogeneities in the nuclues and/or the sporadic nature of outgassing. On the other hand, our high model values show that Borisov could not be made of pure CO, but is more in line with the $\sim 10-20\%$ vs. water maximal abundances of CO found in solar system comets \citep{Bockelee2017}.}

\subsection{Analytic Predictions for 1I/`Oumuamua}

As a baseline estimate for 1I/`Oumuamua, {we assume a bulk composition of CO, which is energetically compatible with the {nongravitational} acceleration and revised {\it Spitzer} limits on the production rate given {the} low activity state outbound at 2 au \citep{trilling2018spitzer, Seligman2021}}.
Equation~\ref{eqn:production} with $\mathcal{H}$ corresponding to a CO composition suggests $Q_{\rm gas}=2.8\times10^{28}$ s$^{-1}$ at perihelion {(the leftmost marker in Figure~\ref{fig:Gq5sig})}, and $Q_{\rm gas}=4.4\times10^{26}$ s$^{-1}$ at $2$ au, in agreement with the \citet{Seligman2021} estimate, as expected.

\begin{figure}
\begin{center}
\includegraphics[width=\linewidth,angle=0]{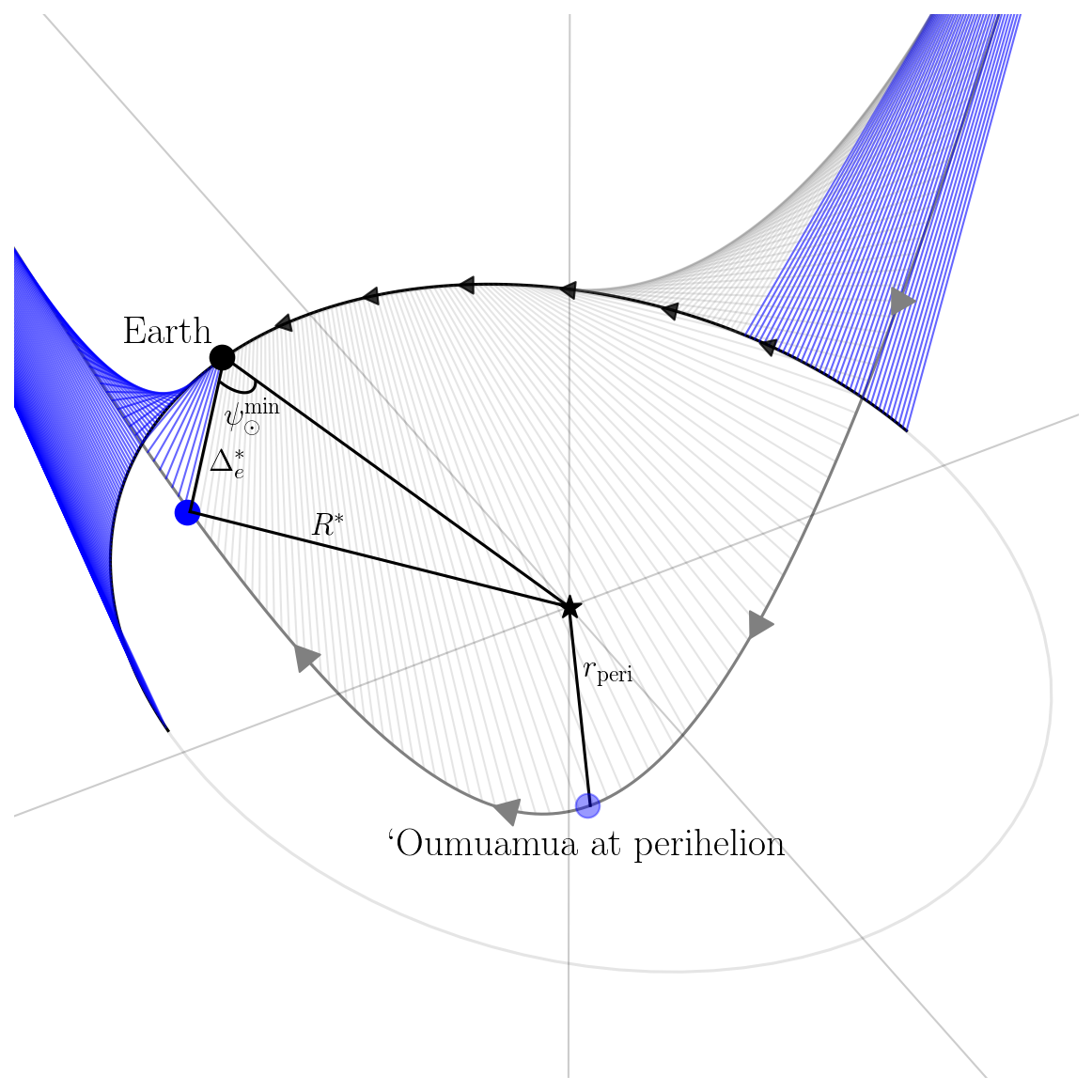}
\caption{{Trajectory and X-ray observability of 1I/`Oumuamua, prior to its discovery. Plotted lines connect the instantaneous positions of Earth and 1I/`Oumuamua. Blue lines denote a solar elongation angle meeting X-ray observing requirements, $\psi_\odot \geq \psi_\odot^{\rm min}$, where $\psi_\odot^{\rm min} = 70^{\circ}$ for {\it XMM-Newton}. The smallest heliocentric distance allowed by this constraint is denoted $R^* = 0.943$ au. At this time, the distance between Earth and 1I/`Oumuamua was $\Delta_e^* = 0.276$ au.}}
\label{fig:oumcomp}
\end{center}
\end{figure}

{With} X-ray flux received at Earth depending strongly on the {target's} distance from the Sun, {1I/`Oumuamua would have only become accessible to {\it XMM-Newton} at {its minimum observable} heliocentric distance $R=0.943$ au. This geometry is depicted in Figure~\ref{fig:oumcomp}. It would have been prudent to observe 1I/`Oumuamua in X-ray at this time, which roughly corresponded with 1I/`Oumuamua's {closest} approach to Earth.}

{
At $R=0.943$ au and $\Delta_e=0.276$ au, we estimate $Q_{\rm gas}=2.0\times10^{27}$ s$^{-1}$, $L_{\rm X} = 1.2\times10^{13}$ erg s$^{-1}$, and {a flux at Earth of} $F_{\rm X} = 5.4 \times 10^{-14}$ erg s$^{-1}$ cm$^{-2}$.} {Assuming an N$_2$ \citep{Desch2021} or H$_2$ \citep{Seligman2020h2} composition yields even higher $F_{\rm X}$, as reported in Table~\ref{tab:lx}.} {For any of these three highly volatile compositions, a} signal would have been unambiguously detected in {a $<1$ ks} X-ray follow-up {exposure}, {producing a unique, easily obtained, positive outgassing detection}.

\subsection{Observations of CX with Comets} \label{sec:cxcomets}

Detections of CX from Solar System comets help gauge conditions under which we may detect CX from ISOs. The detection of X-rays from {C/1996 B2 (Hyakutake)} \citep{Lisse1996} prompted \citet{Dennerl1997} to search archival {\it ROSAT} data for other instances of CX with comet comae. Confirmed cases include: C/1990 Kl (Levy), C/1990 Ni (Tsuchiya-Kiuchi), 45P (Honda-Mrkos-Pajduskova; HMP), and C/1991 A2 (Arai). {The faintest target was 45P, with $F_{\rm X} = 1.5 \times 10^{-13}$ erg s$^{-1}$ cm$^{-2}$, which was detected at high confidence by {\it ROSAT} with a 191 s exposure.}

This comet sample serves as an independent diagnostic for selecting ISOs with detectable CX emission. It is established that a correlation exists between comet X-ray ($L_{\rm X}$) and optical ($L_{\rm opt}$) luminosities \citep[e.g.,][]{Lisse1999b, Lisse2004}. $L_{\rm X}$ plateaus for the brighest comets, but follows linear trend for modest luminosities $L_{\rm X} < 10^{15}$ erg s$^{-1}$ where $L_{\rm X} \sim 10^{-4}L_{\rm opt}$; however, there is dependence on the dust-to-gas ratio [D/G] that induces $1-2$ dex of scatter. Using the trend presented by \citet{Dennerl1997}, we estimate 2I/Borisov's X-ray luminosity as $L_{\rm X}=6.1\times10^{12}$ erg s$^{-1}$ near perihelion, based on its $V=16.6$. For 1I/`Oumuamua, at close approach to Earth and $V=19.7$, we find $L_{\rm X}=2.6\times10^{9}$ erg s$^{-1}$. {(Note that this estimate likely underestimates X-ray emission from 1I/`Oumuamua, and serves only as a foil to other estimates presented in this section.)}

{It is worth mentioning that dusty comae \citep[e.g., C/Hale–Bopp 1995 O1 or 17P/Holmes,][]{Lisse1997, Lisse2013} may destroy solar wind ions without concomitant CX X-ray production \citep{Dennerl1997, Lisse2004}. This effect may prove an obstacle to X-ray observations of dusty ISOs. However, the typical dust content of ISOs is unclear. \citet{Jewitt2017} and \citet{Meech2017} placed stringent upper bounds on 1I/`Oumuamua's dust production rate, at $2\times10^{-4}$ kg s$^{-1}$ and $1.7\times10^{-3}$ kg s$^{-1}$ respectively. Estimates of 2I/Borisov's reach as high as 35 kg s$^{-1}$ \citep{Kim2020}.}

\begin{table*}
\centering
\caption{{Estimated X-ray luminosity ($L_{\rm X}$) and flux ($F_{\rm X}$) of 2I/Borisov and 1I/`Oumuamua {using our model}. For 1I/`Oumuamua, quoted outgassing rates ($Q_{\rm gas}$) depend on the object's assumed bulk composition, and heliocentric ($R$) and geocentric ($\Delta_{e}$) distances. 1I/`Oumuamua's pericenter was $R = r_{\rm peri} = 0.25$ au; however, the solar elongation constraint for {\it XMM-Newton} was first satisfied at $R = 0.943$ au. For 2I/Borisov, we quote values based on our CX emission model, but adopted the measured $Q$(CO) \citep{Bodewits2020}. For both objects we also estimate $L_{\rm X}$ from an empirically derived relationship with $L_{\rm opt}$.}}
\medskip
\setlength{\tabcolsep}{6.0pt}
\begin{tabular}{@{}llcccr@{}}
\hline
Object & Method & $R$ (au) & $Q_{\rm gas}$ (s$^{-1}$)  & $L_{\rm X}$ (ergs s$^{-1}$) & $F_{\rm X}$ (ergs s$^{-1}$ cm$^{-2}$) \\
\hline
2I/Borisov   & Measured $Q$(CO) and modeled CX emission.                     & $2$     & $1.07\times10^{27}$ & $7.5\times10^{11}$ & $6.2\times10^{-17}$ \\
2I/Borisov   & Empirical relationship between $L_{\rm opt}$ and $L_{\rm X}$. & $2$     & -                   & $6.1\times10^{12}$ & $6.0\times10^{-16}$ \\
1I/`Oumuamua & Modeled $Q$(CO) and CX emission.                              & $0.943$ & $2.0\times10^{27}$  & $1.2\times10^{13}$ & $5.4\times10^{-14}$ \\
1I/`Oumuamua & Modeled $Q$(CO) and CX emission at $r_{\rm peri}$.            & $0.25$  & $2.8\times10^{28}$  & $3.3\times10^{16}$ & $7.7\times10^{-12}$ \\
1I/`Oumuamua & Modeled $Q$(N$_2$) and CX emission.                           & $0.943$ & $1.3\times10^{28}$  & $4.7\times10^{14}$ & $2.2\times10^{-12}$ \\
1I/`Oumuamua & Modeled $Q$(H$_2$) and CX emission.                           & $0.943$ & $1.5\times10^{28}$  & $6.9\times10^{14}$ & $3.2\times10^{-12}$ \\
1I/`Oumuamua & Empirical relationship between $L_{\rm opt}$ and $L_{\rm X}$. & $0.943$ & -                   & $2.6\times10^{9}$ & $3.6\times10^{-17}$ \\
\hline
\end{tabular}
\label{tab:lx}
\end{table*}

\section{Future ISO CX X-ray Expectations} \label{sec:expect}

For the prospects of detecting CX X-ray emission with a future ISO, it is important to establish: (1) a set of criteria that provides high-odds of detecting CX X-ray emission; and (2) an estimated fraction of ISOs discovered that will exhibit a detectable X-ray flux. To accomplish these goals, we conduct a Monte Carlo analysis {assuming Rubin Observatory sky parameters \citep[as the largest and deepest continual all-sky survey of the next decade, e.g.,][]{Li2022}} which incorporates baseline predictions for the distribution of perihelia and geocentric distances of ISO trajectories, as well as their size distribution and possible compositions.

\subsection{Trajectories}

{Our Monte Carlo analysis involves randomly drawing orbital configurations for ISOs from previous population synthesis studies. We predict X-ray flux at the minimum heliocentric distance that satisfies the solar elongation angle constraint of {\it XMM-Newton}. This alignment occurs at time $t^*$, defined as:} 
{
\begin{equation}
\begin{aligned}
 t^* \equiv \underset{t}{\operatorname{argmin}} \quad & \|\mathbf{R}(t)\|\\
\textrm{s.t.} \quad & \cos^{-1}\left( -
\frac{\left(\mathbf{R}(t)-\mathbf{R'}(t)\right) \cdot \mathbf{R'}(t)}{\|\mathbf{R}(t)-\mathbf{R'}(t)\|\|\mathbf{R'}(t)\|}
\right)  \geq \psi_{\odot}^{\rm min} \,,
\end{aligned}
\label{eqn:ropt}
\end{equation}}
{where $\mathbf{R}$ and $\mathbf{R'}$ are vectors representing the positions of the ISO and Earth, respectively, with the Sun at the origin. The ISO's hyperbolic trajectory may be expressed in a Cartesian coordinate system where $\mathbf{R} = \{R_1, R_2, 0\}$:}

{
\begin{multline}
    R_2 = \pm \sqrt{e^2-1} \\ 
    \times \sqrt{
    (R_1 + eq/(e-1))^2 - (q/(e-1))^2
    }\,,
\end{multline}}
{with eccentricity $e>1$ and perihelion $r_{\rm peri}$ sampled from distributions provided by \citet{Hoover2022}.
Their study} coupled dynamical simulations with the local stellar velocity dispersion to constrain the population of 1I/`Oumuamua-like ISOs that the {Rubin Observatory} will find. {They} computed the probability density of perihelia, which indicates that $10\%, 25\%$, and $50\%$ of detected ISOs will attain $r_{\rm peri} \leq 0.35$, $0.54$, and $0.83$ au, respectively. Under this distribution, it is clear that 2I/Borisov had an unlikely large perihelion {at 2 au} (and was anomalous with respect to other orbital parameters, such as distance from the solar apex at encounter, and eccentricity). {By contrast}, 1I/`Oumuamua's perihelion {distance of 0.25 au} was unexpectedly {small}.

{Next, we sample Earth's position as a} unit vector with azimuthal and polar angles jointly distributed according to $p(\theta, \phi) = \sin{\phi}/4\pi$, and scaled by {1 au}. In reality, {the orientation of the ISO's trajectory with respect to the ecliptic is} determined by the inbound velocity vector and impact parameter. Stellar kinematics are a good proxy for the galactic distribution of ISOs; however, the typical ages of ISOs are poorly constrained at a population level. We assume the above distribution of ($\theta$, $\phi$) for our order-of-magnitude calculations. {The distances $R^* = \|\mathbf{R}(t^*)\|$ and $\Delta_e^* = \|\mathbf{R}(t^*)-\mathbf{R'}(t^*)\|$ are solved for numerically by fixing Earth's position and assuming $\psi_{\odot}^{\rm min} = 70^{\circ}$. These distances are shown in Figure~\ref{fig:oumcomp}, with 1I/`Oumuamua's trajectory as an example. In cases where the entirety of the ISO's orbit lies within the solar avoidance zone, the Monte Carlo X-ray flux is immediately set to zero.}

We neglect the hyperbolic velocity $v_{\infty}$ in our analysis since the vast majority of ISOs accessible to the {Rubin Observatory} will have $v_{\infty} \leq 40$ \kms. Larger velocities, however, would augment the solar wind ion collision rate by a factor of $10\%$ or greater. It is prudent to conduct X-ray observations of such ISOs during the inbound portion of their trajectories.

\subsection{Bulk Properties}

The composition of a neutral medium does not strongly affect the X-ray power density from CX {in the optically thick regime}. However, the coma's density profile depends on the neutral production rate, which in turn depends on the volatility of surface material in the nucleus of an ISO. While Equation~\ref{eqn:production} estimates neutral production for bulk compositions similar to {C/1996 B2 (Hyakutake)}'s, it is straightforward to account for alternative materials. The flux of molecules ejected from a unit surface area patch of the object (in units of s$^{-1}$ cm$^{-2}$) is \citep[e.g.,][]{Seligman2022},

\begin{equation} \label{eqn:prodcomp}
    \mathcal{N} = \frac{(1-p)I(t) - \epsilon \sigma T_{\rm Sub}^4}{\Delta H/\mathcal{N}_{A}+\gamma k_B\textit{T}_{\rm Sub}}.
\end{equation}
In the above, $p$ is the albedo, $I(t)$ is the time-dependent, solar irradiance, and $\epsilon$ is the surface emissivity. Besides $I(t)$ and physical constants, each parameter in the above equation is material-dependent. In practice, grey-body emission from the minor body is negligible and the incident radiation term dominates. Using the definition in Equation~\ref{eqn:enthalpyrelation}, exchanging $I(t)$ for the incident solar flux, and integrating Equation \ref{eqn:prodcomp} over the illuminated surface area ($\Sigma_{I}$) yields,

\begin{equation} \label{eqn:prodcomp2}
    Q_{\rm gas} = (1-p)\iint_{S}\,\bigg(\,\frac{L_{\odot} }{4\pi R^2\mathcal{H}}\,\bigg)\, d\Sigma_{I}\,.
\end{equation}
Albedo is unconstrained but is likely of order $p \sim 0.1$ for most minor bodies in the Solar System. The average projected surface area  over all viewing angles is $1/4$ times the total surface area for any convex object \citep{Meltzer1949}, so $\Sigma_{I}\simeq\Sigma/4$, where $\Sigma$ is the total surface area of the object. For a sphere, $\Sigma_{I}=\pi a^2$, where $a$ is the radius as defined earlier in this paper.  Therefore, Equation \ref{eqn:prodcomp2} has the same functional form as Equation~\ref{eqn:production}.

\citet{Hoover2022} noted that the size distribution of ISOs could enhance the {Rubin Observatory}'s yield and estimated detection rates for a range of absolute magnitudes (see their Table 2). Many minor body populations adhere to a {power-law} size distribution

\begin{equation} \label{eqn:sizedistr}
    n(>a) = ka^{-\beta},
\end{equation}
for some powerlaw index $\beta$ and normalization constant $k$. \citet{Do2018} note that $\beta > 3$ for ISOs, or else the mass would diverge at large radii. In a collisionally evolved system $\beta = 2.5$ \citep{Dohnanyi1969}. However, Solar System comets have a more complex collisional history, {with} a powerlaw index that depends on the size regime. For example, $\beta = 1.45$ in the $1-10$ km range, and $\beta = 1.91$ in the $2-5$ km subset \citep{Meech2004}. For main-belt asteroids, $\beta = 1.0-1.4$ for sub-km diameters (depending on location within the belt), and $\beta = 1.8$ for larger objects \citep{Yoshida2003}. The distribution for ISOs may be shallow in the case that they are ejected from their respective systems before attaining collisional equilibrium. Also, the bulk of ISOs may be well-described by a single value of $\beta$, but the frequency will necessarily drop off for the largest and smallest objects due to physical limitations. 
{We incorporate the uncertainty surrounding $\beta$ into our analysis by drawing it randomly from a uniform distribution: $\beta \sim \mathcal{U}(1.0, 2.0)$.}
As more ISOs are detected, it will be important to reconsider their size distribution.

\subsection{X-ray Flux}

We further restrict the number of free parameters by assuming the {Rubin Observatory} will detect $1.5$ ISOs per year, the average of the optimistic and conservative rates presented by \citet{Hoover2022}. The remaining free parameters are $a$, {$R$}, and composition. {We randomly draw a nuclear radius $a$ by first drawing a value for $\beta$, and subsequently sampling Equation~\ref{eqn:sizedistr}}. We truncate the distribution at $a_{\rm min}=25$ m and $a_{\rm max}=10$ km, which prohibits ISOs larger than most Solar System comets and also guarantees they are large enough to realistically be detected by the {Rubin Observatory} (an object with $1/4$ the radius of 1I/`Oumuamua and with the same albedo will have $V=24$ at the same $q=0.25$ au). The new cumulative {probability} distribution is:

\begin{equation} \label{eqn:newsizedistr}
    n(>a) = \frac{(a^{-\beta} - a_{\rm max}^{-\beta})}{(a_{\rm min}^{-\beta} - a_{\rm max}^{-\beta})}.
\end{equation}
{The minimum allowed heliocentric distance $R=R^*$ is determined per Equation~\ref{eqn:ropt}}, and the neutral production is calculated using Equation~\ref{eqn:production} for four representative pure compositions: H$_2$, N$_2$, CO, and H$_2$O. {We make an important modification, however, by drawing the heliocentric distance scaling exponent from a random uniform distribution $x \sim \mathcal{U}({-4.33, -1.82})$ as opposed to a fixed $x = -2$, per \citet{Combi2019}.} The CX flux is then calculated using Equation~\ref{eqn:intensity} and an optically thick cross-section given by Equation~\ref{eqn:b} and correction factor $\kappa$.

\begin{figure}
\begin{center}
\includegraphics[width=\linewidth,angle=0]{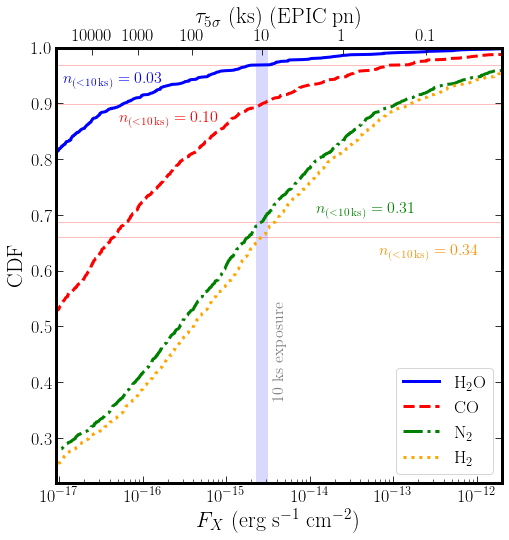}
\caption{Cumulative distribution function of X-ray flux ($F_{\rm X}$) for ISOs detected by the {Rubin Observatory}. Equivalently, the fraction of ISOs detected by the {Rubin Observatory} that will exhibit a given X-ray flux or less. Four distributions are depicted, each corresponding to an ISO population comprised of a different pure volatile: H$_2$O (blue), CO (red), N$_2$ (green), or H$_2$ (orange). On the top x-axis, $F_{\rm X}$ is converted into an exposure duration ($\tau_{5\sigma}$) that would yield a 5$\sigma$ detection with {\it XMM-Newton} EPIC pn. The vertical blue line delineates $F_{\rm X} \simeq 2.7\times 10^{-15}$ erg s$^{-1}$ cm$^{-2}$; a signal which would be detected at 5$\sigma$ confidence with at 10 ks exposure. For example, {about 10\% of CO-dominated} ISOs detected by the {Rubin Observatory} will exhibit this X-ray flux or greater.
}
\label{fig:epicpn5sigma}
\end{center}
\end{figure}

Figure~\ref{fig:epicpn5sigma} summarizes our results, and demonstrates the viability of using X-rays as a diagnostic probe of ISOs across different compositional assumptions. {As discussed earlier, composition has a profound effect on the extent of the outgassed coma and the expected X-ray flux.}
In the most conservative scenario that ISOs are predominantly composed of water ice, then about 3\% found by the {Rubin Observatory} will exhibit $F_{\rm X} \gtrsim 2.7\times 10^{-15}$ erg s$^{-1}$ cm$^{-2}$, which can be robustly detected by a {10 ks} exposure (i.e., $\tau_{5\sigma} = 10$ ks corresponds to a 5$\sigma$ detection; about 6\% are accessible with a $\tau_{5\sigma} = 100$ ks exposure). {Another} {$90\%$} of H$_2$O-dominated ISOs will exhibit undetectable levels of CX emission (i.e., $\tau_{5\sigma}>$1000 ks). If CO is the dominant constituent of ISOs, then about {$10\%$} will emit X-rays detectable with a 10 ks exposure. {The detectability} fraction is even higher for the most volatile compositions, reaching {$31\%$} for N$_2$ ice and {$34\%$} for H$_2$ ice. {With a longer 100 ksec exposure, these fractions reach 19\%, 47\%, and 50\% for CO, N$_2$, and H$_2$, respectively.}

These statistics suggest that {the {Rubin Observatory} will discover new ISOs amenable to X-ray spectroscopic characterization}. If the {survey} finds 15 2I/Borisov analogues (i.e., CO-dominated interstellar comets) over a ten-year campaign, then approximately {$1-2$} of them will be accessible with a 10 ks EPIC pn exposure. The {Rubin Observatory} will probably detect one such ISO in its first {five} years of operation. The approach is similarly effective for 1I/`Oumuamua analogues. X-ray emission should be detectable for {about one third of} N$_2$-dominated and H$_2$-dominated ISOs. On the other hand, consistent non-detections of CX emission will point to an alternative, refractory composition. For objects that exhibit {nongravitational} acceleration, lack of X-rays may also indicate an alternative mechanism that provides the required force (e.g., radiation pressure).

\section{Conclusions} \label{sec:con}

In this paper, we propose X-ray spectroscopy as a means of detecting outgassed comae and revealing {highly volatile} compositions of interstellar objects, a new class of astrophysical phenomena. The {neutral gas} comae of ISOs {should} undergo charge exchange with solar wind ions in the same manner as has been widely observed with Solar System comets. 
We suggest a tiered list of scientific goals that are realistically achievable by existing and concept X-ray missions, sorted by feasibility:

\begin{enumerate}
    \item Photometric detection of CX (i.e., at $\geq 5\sigma$ significance) in order to detect an outgassed coma. Modest exposure times of order 10 ks may reveal a statistically significant X-ray flux originating from an interstellar object, particularly when observed at {heliocentric distances $R \lesssim 1$ au}. The viable parameter space is depicted in Figure~\ref{fig:bq} for pn onboard {\it XMM-Newton}. Confidently detecting X-ray emission over a bandpass of $0.5-0.7$ keV would immediately reveal the existence of a coma undergoing CX with {solar wind} ions. X-ray observations of 1I/`Oumuamua near its {close approach to Earth} would have either confirmed the presence of a coma, or placed {an upper} bound on the production rate to a point that another mechanism would be necessary to explain the anomalous acceleration. 
    \item Joint constraints on $a$ and $Q_{\rm gas}$ in order to infer an ISO's volatile content. Once an ISO's ephemeris is determined, the X-ray flux is degenerate with $a$ and $\mathcal{H}$. By assuming a range of plausible albedos, one can constrain $a$ from an optical light curve \citep[e.g.,][]{Mashchenko2019}. The radius is also constrained by the energetics of {nongravitational} acceleration. For example, \citet{Seligman2020h2} suggested various ices (H$_2$, Ne, N$_2$, Ar, O$_2$, Kr, Xe, CO$_2$, H$_2$O) that could be compatible with 1I/`Oumuamua's outbound acceleration. For a subset of these species (N$_2$, H$_2$, CO, Ne, plus CH$_4$), \citet{Jackson2021} showed the relationship between {nongravitational} acceleration, albedo, and mean spherical radius. Across these different species, the allowable range for $a$ lies between $20-25$ m for albedo $p > 0.6$. However, their respective sublimation energies ($\mathcal{H}$) vary by a factor of $10$, which, according to our model, implies $\mathcal{A}$ varies by a factor of $100$. This interaction area is probed by the X-ray flux from CX. If the nuclear radius of an ISO can be constrained within a factor of {order} unity, then an anomalously high $F_{\rm X}$ would immediately reveal an exotic, highly-volatile composition due to the inverse proportionality between $Q_{\rm gas}$ and $\mathcal{H}$. We reiterate that a non-detection in X-ray would favor low-volatility ices (e.g., H$_2$O, CO$_2$) or an alternative acceleration mechanism.
    \item Measuring line strength ratios towards robustly determining the coma composition. {\citet{Mullen2017} demonstrated that relative CX emission line strengths (including strong transitions from C{\sc v}, C{\sc vi}, and O{\sc vii}) depend on the neutral target medium. Specifically, N$_2$, H$_2$O, O, OH, CO, and CO$_2$ targets were considered in their study. A high-fidelity X-ray spectrum may be achievable by existing facilities for the brightest ISOs. However, this method is most amenable to next-generation X-ray telescopes which feature both large effective areas and high-spectral resolution. A detailed study of this approach, and its feasibility with X-ray concept missions, would be scientifically useful.}
\end{enumerate}

Baseline estimates of the {Rubin Observatory}'s yield suggest around 15 new ISOs over a ten-year campaign. If these objects are predominantly composed of {CO}, about {one or two} will exhibit detectable levels of X-ray flux. {In the case} that most ISOs are comprised of H$_2$ ice, CX will be observable for about {five} of them. Efforts to understand this mysterious population benefit from broadband spectroscopy --- infrared, optical, and UV observations provide complementary insights into physical {and chemical} properties of the nucleus and coma. As X-ray observations of CX with comets have been an active area of study, similar efforts directed toward ISOs will at the very least reaffirm findings made at other wavelengths, and potentially discern the true compositions of these minor bodies.

\section*{Acknowledgments}
We thank the referees of this manuscript for their constructive comments, which significantly improved the presentation of the paper. We extend particular thanks to Carey Lisse for a number of detailed and insightful recommendations which led to a much stronger final version. We also thank Greg Laughlin for his review of the manuscript and for constructive discussions. D.Z.S. acknowledges financial support from the National Science Foundation  Grant No. AST-2107796, NASA Grant No. 80NSSC19K0444 and NASA Contract  NNX17AL71A from the NASA Goddard Spaceflight Center.  D.Z.S. is supported by an NSF Astronomy and Astrophysics Postdoctoral Fellowship under award AST-2202135. This research award is partially funded by a generous gift of Charles Simonyi to the NSF Division of Astronomical Sciences.  The award is made in recognition of significant contributions to Rubin Observatory’s Legacy Survey of Space and Time.

\appendix

\section{Review of Solar Wind Charge Exchange} \label{sec:review}

{Since the discovery of strong X-ray emission from a cool cometary source, C/Hyakutake 1996B2, in 1996 by Lisse et al.,} solar wind charge exchange has gained considerable traction for its ability to explain X-ray emission from objects within the Solar System. CX models have successfully explained observed X-rays from the volatile comae of Solar System comets \citep{Cravens1997, Wegmann1998}, as well as emission originating from Earth \citep{Wargelin2004, Fujimoto2007}, Mars \citep{Kallio1997, Dennerl2002b}, {Venus \citep{Dennerl2002a}}, and Jupiter \citep{Metzger1983, Cravens2003}. 

In order to produce X-ray emission, heavy ions from the solar wind interact with cool atomic and molecular gas surrounding the object. This process also produces a source of X-ray emission from 
{star-forming} galaxies such as M82 \citep[e.g.,][]{Liu2012,Zhang2014}.
{In such a galaxy, CX occurs at the interface between hot and cool interstellar gases, likely responsible for the enhanced forbidden transitions of K$\alpha$ triplet emissions from He-like ions and high-order transitions in Lyman series of H-like ions \citep{Zhang2014}. There has also been tentative detection of charge exchange in galaxy clusters \citep{Gu2018}.}

The general CX reaction is
\begin{equation}
    {\rm A^{q+} + N \xrightarrow[]{} (A^*)^{(q-1)+} + N^+},
\end{equation}
where the ion ${\rm A^{q+}}$ gains an electron from the neutral species ${\rm N}$ (atomic or molecular, often H, H$_2$ and He). The integer ${\rm q}$ denotes the initial ionization state of the species A. The electron is in an excited state, denoted by the superscript $*$.
{As it cascades to the ground state, the electron emits X-ray and UV photons.}
Implicit in this equation are initial and final state quantum numbers ($n$, $l$, and $m$) of the transferred electron. The most probable final-state principal quantum number $n$ is given by Equation 2.6 of \citet[][]{Janev1985}, which is an approximation that is generally applied while modeling this process \citep{Smith2012}. 

The CX interaction cross-section ($\sigma_{\rm sq}$) for a given species (denoted by subscript s) and charge (denoted by subscript q) was modeled by \citet{Wegmann1998}, and is given by, 
\begin{equation}
\label{eqn:cs}
    \sigma_{\rm s q} = \,\bigg(\, \frac{{\rm q}-1}{{\rm q}^2/2n^2 - |I_{{\rm p, s}}/27.2 \, {\rm eV}|}\,\bigg)^2\,\times0.88\times10^{-16} \; ({\rm cm^2}).
\end{equation}
The variables ${\rm q}$ and $I_{{\rm p, s}}$ in Equation \ref{eqn:cs} correspond to the dimensionless integer charge of the ion and ionization potential (measured in eV) of the target species, respectively. 

For comets, the solar wind typically interacts with volatile coma particles. Some of the most common target species outgassed by Solar System comets are water and its constituents (O, OH, H) \citep{Cochran2015,Biver2016,Bockelee2017}. All of these species have similar ionization energies $\sim13$ eV \citep{Wegmann1998}. CX cross-sections are $\sim 10^{-15}$ to $10^{-14}$ ${\rm cm^2}$ for heavy solar wind ions such as O\textsc{viii}, O\textsc{vii}, C\textsc{vii}, C\textsc{vi}, N\textsc{vii}, Ne\textsc{ix}, Si\textsc{x}, and Fe\textsc{xii}. That is, charge exchange occurs when these ions come within $\sim 1$ nm of the neutral targets. The energy difference between the captured electron's initial (excited) state and ground state ($E_{\rm excit}$) can reach several hundred eV. Therefore, these transitions provide significant contributions to EUV and X-ray radiation. {CX emission from cometary comae has rich observational and theoretical backing, as reviewed by \citet{Bodewits2012} and \citet{Dennerl2012}.}
 
Our main concern is charge exchange between the solar wind and the outgassed species within an ISO's coma. Assuming that a given solar wind ion undergoes charge exchange once during its passage through the neutral medium, the X-ray power density $P_{\rm sj}$ for transition (j) at a given position ${\mathbf r}$ is:

\begin{equation} \label{eqn:power}
    P_{\rm sj}({\mathbf r}) = \Phi_{\rm sq}({\mathbf r})\,\sigma_{\rm sq}\,b_{\rm sqj}\,n_{\rm n}({\mathbf r})\,\Delta E_{\rm sqj}\,,
\end{equation}
\citep{Cravens1997, Cravens2000a, Cravens2009} which has units of eV cm$^{-3}$ s$^{-1}$.
Here $\Phi_{\rm sq}$ denotes the solar wind flux of a given ion, $\sigma_{\rm sq}$ is the CX cross-section (Equation~\ref{eqn:cs}), $b_{\rm sqj}$ is the spectral cascading probability for transition (j) which releases a photon of energy $\Delta E_{\rm sqj}$, and $n_{\rm n}$ is the number density of neutral species. The solar wind flux may be written as $\Phi_{\rm sq}=f_{\rm sq}n_{\rm SW} u_{\rm SW}$, where $n_{\rm SW}$ and $u_{\rm SW}$ correspond to solar wind proton number density and velocity, respectively. The solar wind fraction of a given species/charge is $f_{\rm sq}$. These quantities are functions of position in space. For example, the unshocked solar wind has a typical speed of $\sim400$ \kms\ and density of $\sim0.4$ cm$^{-3}$ at 5 au \citep{Cravens2003}. The density $n_{\rm SW}$ falls off as the square of distance from the host star \citep{Cravens2000a}. While the solar wind flux varies over time, the fraction of heavy ions to protons in the solar wind remains fairly consistent at $f\approx10^{-3}$. Intensity is obtained by integrating Equation~\ref{eqn:power} over a given path length (i.e., performing a line integral). 

\section{Charge Exchange with Outgassed Comae} \label{sec:coma}

{Our prescription for CX emission is calibrated to observations of C/1996 B2 (Hyakutake) and early models that described its X-ray luminosity.}
\citet{Wegmann1998} calculated the total X-ray emissivity  by summing over all solar wind ions in Equation~\ref{eqn:power}. They adopted parameter values of $\sigma_{\rm sq} = 3\times10^{-15}$ cm$^{2}$, $f_{\rm O} = 0.0005$, and $1100$ eV emitted per oxygen ion in the solar wind based on summing the CX excited state energies of all ions and weighting them according to their relative abundance fraction. Further, they assumed an `effectivity' of $0.4$ which replaces the $b_{\rm sqj}$ terms. The power density is approximately

\begin{equation}
P_{\rm X}({\mathbf r'}) = 4\times10^{-20}n_{\rm SW}n_{\rm n}({\mathbf r'}) {\rm \, erg \, cm^{3} \, s^{-1}} \, , 
\end{equation}
where the density of neutrals at distance $\mathbf{r'}$ from the comet's nucleus follows 

\begin{equation} \label{eqn:coma}
    n_{\rm n}(\mathbf{r'}) = \,\bigg(\, \frac{Q_{\rm gas}}{4\pi v \mathbf{|r'|}^2}\,\bigg)\,e^{-{|\mathbf{r'}|}/{\lambda}}\,,
\end{equation}
for a total neutral production rate $Q_{\rm gas}$, outflow velocity $v$, and {photodestruction} length scale $\lambda$. \citet{Wegmann1998} adopted
$Q_{\rm gas}=1.5\times10^{29}$ s$^{-1}$ and $n_{\rm SW} = 7$ cm$^{-3}$. {By neglecting photodestruction, which \citet{Cravens1997} deemed negligible within $5\times10^5$ km, and adopting} a nominal $v=1$ \kms, integration over a sphere of beam radius $135,000$ km 
yields $L_{\rm X} \sim 6\times10^{15}$ erg s$^{-1}$, comparable to the measured luminosity and other estimates in the literature. As follows, we determine expectations for CX emission from the comae of other minor bodies.

\subsection{Luminosity from the Optically Thick Coma}

\begin{figure}
\begin{center}
\includegraphics[width=0.5\linewidth,angle=0]{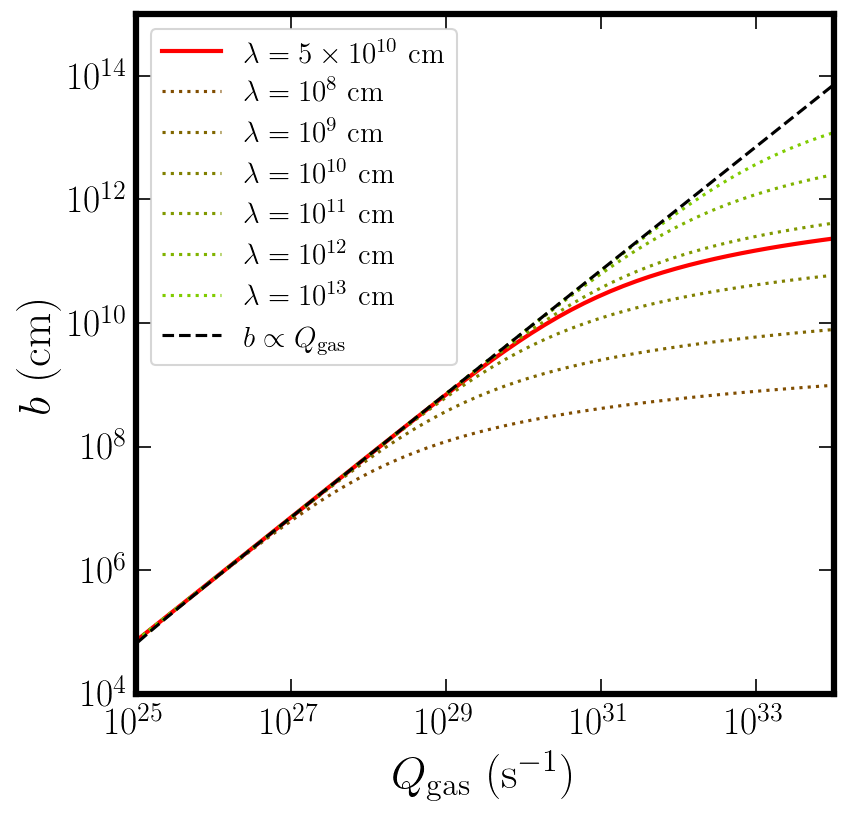}
\caption{Relationship between outgassing rate $Q_{\rm gas}$ and radius $b$ within which a coma is optically thick to CX. Since $b$ lacks a closed form (Equation~\ref{eqn:b}), it is solved for numerically. The function is plotted for several assumptions of the ionization scale $\lambda$. In all cases $v=1$ \kms. Our main analysis assumes $\lambda = 5\times10^{10}$ cm (red line). As $\lambda$ increases, the function approaches a linear relationship between $Q_{\rm gas}$ and $b$ (black dashed line).
}
\label{fig:bq}
\end{center}
\end{figure}

CX produces a {surface luminosity}, $4\pi I$ \citep{Cravens2003}, given by the equation,

\begin{equation} \label{eqn:intensity}
    4 \pi I = n_{\rm SW} \, u_{\rm SW}f\,N\,\Delta E\,.
\end{equation}
Each ion contributes, on average, $N$ photons of typical energy $\Delta E$, and the solar wind {minor} ion fraction is given by $f$. The {total} luminosity is obtained by integrating Equation \ref{eqn:intensity} over the effective surface area. Our baseline model assumes that this area is the region where the coma is optically thick to CX. That is, it satisfies the criterion:

\begin{equation}\label{eqn:opt_thick}
    \int_{S} \, (\,\sigma_{\rm sq} n_{\rm n}({\mathbf r'}) \,)\, ds > 1
\end{equation}
for a linear
projected path $S$ {of the solar wind through the coma}, where ${\mathbf r'}$ is the distance from the cometary nucleus as in Equation \ref{eqn:coma}. {As long as the coma neutral column is large enough that the solar wind is fully depleted of all highly stripped minor ions, the individual transition
probabilities are not important. For simplicity, we also fix the cross-section at $\sigma_{\rm sq} = 3\times10^{-15}$ cm$^{2}$. This value is consistent with expectations from the ionization potentials of H$_2$ ($I_{p,s} \approx 15.4$ eV) and N$_2$ ($I_{p,s} \approx 15.6$ eV). Considering these compositions and those more typical of comets, 
 we find $2\times10^{-15}$ cm$^{2}$ $\leq \sigma_{\rm sq} \leq 13\times10^{-15}$ cm$^{2}$ for ion charges $+5 \leq q \leq +7$.}
We assume that the coma is approximately spherical and follows a neutral density profile given by Equation~\ref{eqn:coma}. We define the impact parameter, $b$, as the minimum distance between the path $S$ and the nucleus. Also, let $R$ be the distance between the Sun and the nucleus, and assume that $R \gg b$. We parameterize the path $S$ as a function of $\mathbf{r'}$. We define the dimensionless parameter $x\equiv|\mathbf{r'}|/b$ and a coefficient $K\equiv \sigma_{\rm sq} Q_{\rm gas}/4\pi w$. By substituting Equation \ref{eqn:coma} into Equation \ref{eqn:opt_thick} with these new parameters, the optically thick transition occurs where

\begin{equation}
    \frac{2K}{b}\int_{1}^{\infty} \frac{1}{x\sqrt{x^2-1}}e^{-bx/\lambda} dx = 1 \,,
\end{equation}
or, equivalently, where

\begin{equation} \label{eqn:b}
\frac{2K}{b}\int_{0}^{\pi/2} e^{-b\, {\rm sec}{\theta}/\lambda} d\theta = 1 \,.
\end{equation}
Equation~\ref{eqn:b} is equivalent to a function $b=b(\sigma_{\rm sq}, w, \lambda, Q_{\rm gas})$ which may be evaluated numerically. It returns a distance within which the coma is optically thick to CX. For physically plausible {photodestruction} scales of $\lambda = 10^{8}-10^{11}$ cm \citep{Combi2004},
a reasonable approximation is that $b \propto Q_{\rm gas}$ in the regime $Q_{\rm gas} < 10^{28}$ s$^{-1}$. Figure~\ref{fig:bq} shows the validity of this approximation for different assumed $\lambda$.
As $\lambda \rightarrow \infty$, the function approaches a linear relationship.

Taking {C/1996 B2 (Hyakutake)} as a nominal example {(hence the above assumed $\lambda$ and $Q_{\rm gas}$ values)},
then we find $b \approx 10,000$ km which is in reasonable agreement with the 30,000 km penetration depth estimated by \citet{Wegmann1998}. In order to estimate total X-ray luminosity, we assume $n_{\rm SW} = 7$ cm$^{-3}$ at 1 au, $u_{\rm SW} = 400$ km s$^{-1}$, $f=0.001$, $N=1$, and $\Delta E = 550$ eV. The above equations predict $L_{\rm X} = 8.2\times10^{14}$ erg s$^{-1}$ from the optically thick region alone. This estimate is about $5\times$ less than the total luminosity measured by {\it ROSAT}. Using a refined hydrodynamical model, \citet{Wegmann2004} also estimated $L_{\rm X}$ for {C/1996 B2 (Hyakutake)} during its close approach to Earth. Their measurements were centered around $10^{16}$ erg s$^{-1}$, which is higher than the initial estimate from \citet{Lisse1996}. The discrepancy between these literature estimates and our own is probably due to significant emission that took place in optically thin regions of the coma, currently unaccounted for in our model for the optically thick zone. Therefore, it is useful to calibrate our model and correct for this component per the following.

\subsection{Extensions to the Analytic Model}

Holding other parameters constant, $b$ scales linearly with $Q_{\rm gas}$ in the regime $Q_{\rm gas} < 10^{30}$ s$^{-1}$ for $\lambda \sim 5\times10^{10}$ cm. The CX interaction area in this regime can be approximated by $\pi b^2 \approx \Gamma Q_{\rm gas}^2$, with $\Gamma = 1.5\times10^{-40}$ s$^2$ cm$^2$. A similar conclusion was reached by \citet{Wegmann2004}, who developed a hydrodynamical model of X-ray emission morphology and applied it to observations of comets. They derived a relationship $L_{\rm X} = C H Q_{\rm gas}^2$ where $H = u_{\rm SW} n_{\rm SW} f N \Delta E$ is the heavy ion solar wind flux. By fitting their estimates of $H Q_{\rm gas}^2$ to observed X-ray luminosities of comets, they determined a constant of proportionality $C = 10^{-38}$ s$^2$ cm$^2$. Let $\mathcal{A}$ define an effective, optically thick surface area that would yield a given $L_{\rm X}$. Then the relationship found by \citet{Wegmann2004} is equivalent to $\mathcal{A} = C Q_{\rm gas}^2$. We adopt a correction factor $\kappa \equiv \sqrt{C/\Gamma} \approx 8.1$ that defines a new effective radius,

\begin{equation} \label{eqn:newb}
    \tilde{b} \equiv \kappa b
\end{equation}
In other words, $\kappa$ is a factor that increases the radius of the CX interaction area from $b$ to $\tilde{b}$, thereby approximately accounting for emission from regions beyond the optically thick zone. {Prior observations of CX within cometary comae \citep[e.g.][]{Lisse1999b, Lisse2004} have produced detections of significant X-ray emission extending {as little as} $\sim 5\times10^{4}$ km (for 2P/Encke 2003), {and as far as} an apparent limit at $\sim 10^{6}$ km (for C/1996 B2 (Hyakutake), C/1991 K1 (Levy), and C/1995 O1 (Hale-Bopp)).}

{We adopt} $Q_{\rm gas} \propto a^{2}R^{-2}$, where $a$ is body's radius, and $R$ is the instantaneous heliocentric distance. The $R^{-2}$ dependency is the same as that of the incident solar radiation. {While physically motivated, this scaling simplifies true cometary activity. Following a long-baseline survey of water outgassed by 61 comets, \citet{Combi2019} found generally steeper scaling: $Q({\rm H_2O}) \propto R^{x}$, with $-4.33 \leq x \leq -1.82$ depending on taxonomic class. This finding is incorporated into our expectations for future ISO observations in \S\ref{sec:expect}.} 

Our model must also account for the fact that more volatile species will sublimate and outgas at higher rates (potential compositions are discussed further in \S\ref{sec:expect}). We use the following scaling relation,

\begin{equation} \label{eqn:production}
    Q_{\rm gas} = 1.5\times10^{29} \times \Big(\frac{a}{2.4\,{\rm km}}\Big)^2 \Big(\frac{R}{1\,{\rm \, au}}\Big)^{-2} \Big(\frac{\mathcal{H}}{9.3\times10^{-23} \, {\rm kJ}}\Big)^{-1} {\rm s}^{-1} \,,
\end{equation}
{which is calibrated to the nuclear radius of C/1996 B2 (Hyakutake) \citep{Lisse1999a}, and the production rate adopted by \citet{Wegmann1998}. Production rates for OH were measured by \citet{Schleicher1996}, and varied from $\log_{10} Q({\rm OH}) = 28.89$ (at $R=0.94$ au) to $\log_{10} Q({\rm OH}) = 29.17$ (at $R=1.08$ au). In the above,} $\mathcal{H}$ is the total energy input for each coma particle,

\begin{equation} \label{eqn:enthalpyrelation}
    \mathcal{H} \equiv \Delta H/\mathcal{N}_{A}+\gamma k\textit{T}_{\rm Sub}\, .
\end{equation}
The above expression involves enthalpy of sublimation $(\Delta H)$, temperature of sublimation ($T_{\rm Sub}$), and adiabatic index of the escaping vapor ($\gamma$), all of which are material dependent (Table~\ref{tab:chem}). The quantity $\mathcal{H}$ is the sum of two components: the energy required to sublimate an ice molecule (the first term) and the energy required {to} heat the molecule to the outflow velocity (the second term). In the case of water, $\mathcal{H_{\rm H_2O}} = 9.3\times10^{-23}$ kJ (values for other species are listed in Table~\ref{tab:chem}). Chemical properties also determine the outgassing velocity $v$:
\begin{equation} \label{eqn:wchem}
    v \simeq c_s = \sqrt{\gamma k\textit{T}_{\rm Sub}/\mu m_H} \, ,
\end{equation}
for mean molecular weight $\mu$ and sound speed $c_s$. {For typical values $v = 0.3-1$ \kms, $Q < 10^{30}$ s$^{-1}$, photodestruction timescales $t_{\rm photo} < 10^7$ s, and $\lambda = wt_{\rm photo}$, the optically thick radius does not exceed $10^6$ km under our model. It is limited by the exponential decay term in Equation~\ref{eqn:coma}.} {We reiterate that only monotomic and homonuclear diatomic species (having zero diplole moment) will be more amenable to CX X-ray characterization than to infrared fluorescence spectroscopy.}

The normalization in Equation~\ref{eqn:production} is to {C/1996 B2 (Hyakutake)}, and the expression should be considered an order-of-magnitude estimate of $Q_{\rm gas}$ for other objects. Ideally, one may use independent estimates of $Q_{\rm gas}$ obtained either from infrared spectroscopy or a more detailed theoretical model that accounts for the target's composition. For most Solar System comets, $Q_{\rm gas} \simeq Q({\rm H_2O}) + Q({\rm CO_2}) + Q({\rm CO})$, which is the sum of production rates of the most common volatile components in comets \citep{Bockelee2017}. Nevertheless, Equation~\ref{eqn:production} enables baseline predictions for new minor bodies, and is especially useful for population-level predictions when given distributions for $a$ and perihelion $r_{\rm peri}$. Additionally, we assume that the solar wind proton number density is given by, 

\begin{equation} \label{eqn:nsw}
    n_{\rm SW}(R) = 7\,{\rm cm}^{-3} \, \Big(\frac{R}{1{\rm \, au}}\Big)^{-2},
\end{equation}
which is maximized at perihelion when $R = r_{\rm peri}$. {Note, flares can enhance the solar wind ion flux, thus improving the feasibility of CX X-ray observations at larger $R$. For example, a flare nearly doubled the soft X-ray count rate from C/1999 S4 (LINEAR) \citep{Lisse2001}.} 

Combining Equations~\ref{eqn:intensity} \& \ref{eqn:newb}, we arrive at a general model for X-ray luminosity

\begin{equation} \label{eqn:lxest}
    L_{\rm X} = \pi \tilde{b}^2 n_{\rm SW} u_{\rm SW}fN\Delta E,
\end{equation}
which is a function of $\{\sigma_{\rm sq}, \lambda, a, r_{\rm peri}, \mu, \Delta H, \gamma, T_{\rm Sub}\}$, assuming observations take place at perihelion. Also, a nominal $\lambda = 5\times10^{10}$ cm is adopted for our analysis. We use this model to predict the luminosity of CX emission with ISOs in \S\ref{sec:expect}, and a schematic of the entire CX process with ISOs is shown in Figure~\ref{fig:schematic}. It is worth explicitly highlighting our model's dependence on perihelion distance. We demonstrated an approximately linear relationship between $b$ and $Q_{\rm gas}$, where $Q_{\rm gas} \propto r_{\rm peri}^{-2}$ through its dependence on the solar radiation flux $F_{\odot} \propto r_{\rm peri}^{-2}$. The CX surface area follows $\mathcal{A} \propto b^2$, and the incident heavy ion flux follows $n_{\rm SW} \propto r_{\rm peri}^{-2}$. Therefore, $L_{\rm X} \propto r_{\rm peri}^{-6}$: the perihelion of an ISO strongly dictates whether its CX X-ray emission is detectable {\citep[similar scaling relationships were explored by][]{Lisse1999b}}.

\bibliographystyle{aasjournal} 
\bibliography{main} 

\end{document}